# Association between centrality and flourishing trait: analyzing student co-occurrence networks drawn from dining activities

Yi Cao[1,2], Shimin Cai[2], Xiaorong Shen[3,*], Tao Zhou[1,2,*]

[1]CompleX Lab, University of Electronic Science and Technology of China, Chengdu 611731, People's Republic of China
[2]Big Data Research Center, University of Electronic Science and Technology of China, Chengdu 611731, People's Republic of China
[3]i-Large Model Innovation Lab of Ideological and Political Science, University of Electronic Science and Technology of China, Chengdu 611731, People's Republic of China
* Correspondence to sxr63@uestc.edu.cn, zhutou@ustc.edu

Abstract: Comprehending the association between social capabilities and individual psychological traits is paramount for educational administrators. Presently, many studies heavily depend on online questionnaires and self-reported data, while analysis of the connection between offline social networks and mental health status remains scarce. By leveraging a public dataset encompassing on-campus dining activities over 21 weeks, we establish student co-occurrence networks and closely observe the changes in network topology over time. Empirical analysis shows that the node centralities of the student co-occurrence networks exhibit significantly positive correlation with the enhancement of the flourishing trait within the field of mental well-being. Our findings offer potential guidance for assisting students in maintaining a positive mental health status.

## 1. Introduction

Understanding the relationship between social capabilities and individual psychological traits is an important part of the educational landscape [1-4]. Such understanding can serve as valuable evidence and reference for educational administrators when designing interventions or incentives. In educational settings, the relationship between social connections and individual personality is complicated. In some scenarios, the relationship is positive. For instance, McLoughlin *et al.* [5] conduct a study among students from eight secondary schools in South Australia and reveal that adolescents with stronger social ties exhibit a higher likelihood of responding positively to frequent instances of cyberbullying than those who are not. In addition, Spilt *et al.* [6] and Zheng [7] focus on classroom interactions between teachers and students and unpack that fostering positive interactions not only enhances the well-being of both teachers and students but also yields long-term benefits, including improved student achievement and a heightened sense of belonging for teachers. Analogously, Frenzel *et al.* [8] examine the classroom climate and the self-reported enjoyment of seventh- and eighth-grade students in mathematics and disclose a significant correlation between students' enjoyment of the classroom and teachers' enthusiasm for the lessons. Bond et al [9] find in an analysis of eighth-grade students showing that those with weaker school connections face a 1.3% increase risk of anxiety or depression compared to the control group. Moreover, they exhibit higher probabilities, at 1.7%, 2.0%, and 2.0%, of engaging in alcohol, tobacco, and drug use, respectively. In some other circumstances, however, social relationships and psychological traits show negative associations. For example, Waytz *et al.* [10] conduct four survey studies at the University of Chicago and reveal that excessive consideration of close individuals increases one's tendency to dehumanize others when perceiving them. Through investigating 70 undergraduates, Kalpidou *et al.* [11] show that the number of Facebook friends is correlated with low self-esteem among freshmen and impedes their academic adjustments.

Social relationship data in relevant literature is predominantly derived from online questionnaires, online climbing and self-reported surveys, whereas studies on offline social relationships among students or teachers are limited. Nevertheless, the conclusions from online social networks are inappropriate to directly transplant to offline social networks, because online and offline social networks usually have patent differences in network structure [12]. Furthermore, the social relationships in relevant literature are often constrained by narrow and static data, limiting the ability to explore the intricate connections between social relationships and psychological traits at a more detailed and nuanced level. Hence, more detailed data and more comprehensive and profound investigations are warranted to delve into the intricate relationship between social interactions and individual psychological traits within the campus environment.

Recently, with the increment of data availability and abundance, empirical analyses based on large-scale data are rapidly becoming mainstream to unravel the intricate connections between individual micro-level behaviors and the broader macro-level dynamics of the social economy [13-15]. Although acquiring direct information about offline social relationships between individuals remains challenging, the widespread availability of network communication and sensor technologies in today's world has significantly facilitated the collection and preservation of individuals' daily behaviors. Scientists now have unprecedented opportunities to extract underlying behavioral patterns from the collected daily behaviors and then extend the findings to uncover social relationships or networks among individuals. They can further explore the correlations between these patterns and the attributes of individuals such as psychological traits and socioeconomic status. For example, within a series of studies centered on campus behaviors and academic performance, Cao *et al.* [16] propose an entropy-based indicator to describe and quantify the regularity of students' daily lives in terms of on-campus consumption via smart cards. They highlight that regular daily behaviors have independent effects on academic performance even in the presence of study diligence.

Yao *et al*. [17] further discover a noteworthy association that students who exhibit higher similarity in their daily behaviors also tend to have closer academic outcomes.

The StudentLife project [18], conducted at Dartmouth College, stands out as one of the most renowned endeavors aimed at gathering and analyzing information about students' daily behaviors. The project utilizes sensors on uniformly configured Android phones to automatically collect and store behavioral and location data from 48 students on campus. The finding reveals a strong correlation between these objective sensor data and students' psychological questionnaire responses and academic performance. Inspired by the project's idea of utilizing sensors for automated behavioral data collection, Sano *et al*. [19] investigate a sample of 66 students over 30 days leveraging mobile phones and wearables to gather objective physiological markers. The result indicates that sensor data exhibits significant efficacy in accurately classifying various traits, including academic performance, sleep quality, perceived stress levels, and more. A subsequent observational study by the same research group [20] arrives at similar conclusions with a larger sample size of 201 students. The study reveals that wearables outperform mobile phones in classifying the aforementioned attributes. Nevertheless, the social networks for the above works are derived from self-reports, which may be susceptible to social desirability bias [21,22]. To address the concern of social desirability bias, Liu *et al*. [23] conduct a study at the University of Notre Dame by constructing social networks directly from text message data obtained from approximately 700 undergraduate students, revealing that dynamic networks, as opposed to static networks, can be utilized to provide more accurate predictions on participants' mental health.

Flourishing is one of the most important and promising concepts in positive psychology. The flourishing trait comprehensively reflects one's degree of self-perception in various dimensions, including interpersonal relationships, self-esteem, goal pursuit, and optimism [24]. Investigating the relationship between the flourishing trait and social capabilities holds tremendous significance to the long-term development of individuals. Gudka *et al*. [25] propose a conceptual framework for achieving flourishing through social media, which believes that the use of social media can improve an individual's flourishing trait from four dimensions: obtaining social support, promoting identity, improving life satisfaction, and attaining self-efficacy. Marciano *et al*. [26] suggest that positive online social relationships and engaging social media content play a significant role in promoting the flourishing trait among adolescents. Rahe *et al*. [27] survey 138 adults and find that pro-society is a significant positive predictor of flourishing. Eraslan-Capan [28] analyzes questionnaire data collected from 260 college students and discovers that individuals with sparser social connections are more likely to experience feelings of despair, and their levels of flourishing are correspondingly low.

In educational settings, it remains unclear whether the extent of self-perceived flourishing traits would gradually accumulate and strengthen through individuals' offline interactions with others in their unobtrusive daily activities. Leveraging longitudinal dining check-in data of students on a 21-week time scale provided by the StudentLife project, this study establishes student co-occurrence networks based on dining activities and observes the changes in the topological structure over time. It reveals that the sizes and clustering coefficients of the student co-occurrence networks are related to interpersonal interactions among students during different periods. This study further utilizes the questionnaire data of flourishing scale from the StudentLife project to explore whether dining is associated with such psychological trait. The result shows that there is a significant and positive correlation between the node centrality in the student co-occurrence networks and the enhancement of the flourishing trait.

The rest of this paper is organized as follows. In the next section, we present the data details, including the dining data and the flourishing scale. In Section 3, we introduce the establishment of the student co-occurrence networks and the temporal changes of the topology. The association analysis between the co-occurrence networks and the flourishing trait is depicted in Section 4. Finally, we conclude the study and outline the limitation of this paper.

## 2. Data

The StudentLife project [18] was launched in 2013 in the Department of Computer Science at Dartmouth College. All students who participated in the project enrolled in the CS65 Smartphone Programming class (CS65 course for short in the rest part of this paper) at Dartmouth College in the spring of 2013. A total of 75 students from different grades and backgrounds have enrolled in the CS65 course, 60 of whom participated in the StudentLife project. As the project progressed, seven students withdrew from the project midway, and an additional five students gave up their credits of the CS65 course. As a result, the number of effective participants for the StudentLife project is 48. In the original dataset of the 48 students, all student identification information, including but not limited to names, student IDs, genders, apartment addresses, grades, majors, etc., has been encrypted or erased. It ensures that no specific student can be identified from the existing data.

In the raw data of the 48 project participants, 31 of whom have dining records. The dataset encompasses dining data for the 31 students throughout the spring of 2013, spanning a total period of 21 weeks (2013.01.06-2013.06.02). The dining data consists of 7,482 digital records, which include check-in times for meals and tea breaks, captured at 7 dining establishments on campus. Note that the 11th week corresponds to a holiday period for Dartmouth College, resulting in very few dining records. Consequently, for the rest of this study, week 11 will be excluded, and only the remaining 20 weeks will be considered. When referring to weeks 11-20 in the following context, they actually correspond to weeks 12-21 in the original dataset. The CS65 course is offered during weeks 11-20, during which there are class sessions and psychological questionnaire data available. For weeks 1-10, there is only dining data available.

The 31 students also completed questionnaires about the flourishing trait. The flourishing scale [29] consists of 8 sub-questions, each of which is evaluated on a scale ranging from 1 to 7, resulting in an overall score that spans from 8 to 56. The higher an individual score, the better the self-perception across a series of positive psychology dimensions, including optimism, interaction engagement, relationship development, goal pursuit, and feelings of accomplishment. The questionnaire survey was conducted twice: once on the day before the 11th week began, and another on the day after the 20th week. The content of the two questionnaires is entirely consistent. The score difference in the two surveys reflects the change in the flourishing trait during these 10 weeks.

## 3. Topology analysis

Dining is a high-frequency daily behavior. The timing, venues, and companions associated with dining events inherently contain information about individuals' social preferences. The purpose of constructing the student co-occurrence network is to deduce offline social relationships among students based on dining, one of the most fundamental and representative daily behaviors.

In the StudentLife dataset, the dining information is essentially a series of time-spatial records in the form of a tuple denoted by <timestamp, location>. The timestamp $t_i$ records the precise time of the $i$th dining activity, with time resolution in seconds.

Note that no interaction information of individuals is documented in the raw dining data. Therefore, it becomes essential to establish a criterion to determine co-occurrences among students during dining activities. To address this, we introduce a time threshold $T$. We assume that if the interval of order times between any two students $u$ and $v$ is not greater than $T$ at the same dining location on the same day, it signifies a dining co-occurrence between them. Within the StudentLife dataset, we observe that the on-campus dining establishments primarily consist of convenience stores and coffee bars. In addition, there is no evidence that dining locations for long single-meal times, such as buffets, exist in the dataset. Hence, considering that American-style on-campus meals are generally simple meals, we set the time threshold $T$ as 20 minutes. Formally, the dining-based student co-occurrence is represented as follows:

$$co(u,v) = \begin{cases} 1, & \exists l, i, j, \ s.t. \ \left|t_i^{u,l} - t_j^{v,l}\right| \leq T \\ 0, & \text{otherwise} \end{cases}, \quad (1)$$

where $co(u,v)$ represents whether student $u$ and $v$ have co-occurrence on dining, and $t_i^{u,l}$ means the $i$th dining time of student $u$ at dining establishment $l$.

According to the aforementioned definition of student co-occurrences based on dining activities, we set that within a certain period (which can be a day, a week, or a semester), if there is at least one instance of dining co-occurrence between student $u$ and $v$, a link is formed between them, denoted as $e(u,v) = 1$. Conversely, if there is no dining co-occurrence, we set $e(u,v) = 0$. Considering the typical periodicity of students' on-campus schedules (both learning and living), we weekly construct a student co-occurrence network based on dining records generated within that week. A total of 20 student co-occurrence networks are thus established, denoting as $G_1, G_2, \cdots, G_{20}$.

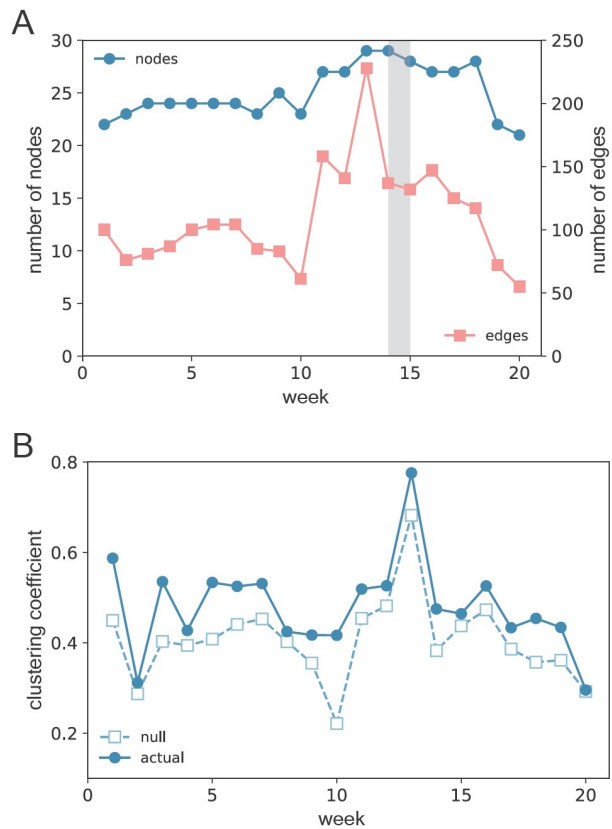

Figure 1. The topology changes of the student co-occurrence networks $G$.. (A) Network size, including both the number of nodes and the number of edges. (B) Clustering coefficients.

Here the student co-occurrence networks $G$. are undirected. Figure 1A illustrates the temporal changes of both the number of nodes and the number of edges within the student co-occurrence networks, where students without any connected edges (isolated nodes) are not considered to be in the co-occurrence networks. Over two-thirds of the nodes (students) have joined the student co-occurrence network in the initial week. Throughout the entire duration of 20 weeks, the number of nodes in the networks consistently remains at a scale no less than this initial level. Around the 13th week, the number of nodes in the student co-occurrence networks reaches its peak at 29, indicating that nearly all individuals have become part of the student co-occurrence network $G_{20}$. The number of edges in the student co-occurrence networks is relatively stable in the first 10 weeks, fluctuating in the range of 60-100. It has a remarkable burst that peaks at 228 during the 13th week. However, this surge is swiftly followed by a sharp decrease to around 140 by the 14th week. The number of edges stabilizes at this level till the 18th week. During the final two weeks of the StudentLife project, there is a notable decrease in the scale of the student co-occurrence networks. The number of nodes decreases to approximately two-thirds of the total number of students, while the number of edges reduces to just over 50 simultaneously. As the 2013 spring semester at Dartmouth College nears its end, it is reasonable to speculate that the observed decline in the scale of the student co-occurrence networks is correlated with the various end-term tests. During this period, students are likely to be occupied with intensive revision either in dormitories or in the library, which could explain the reduced level of connectivity within the student co-occurrence networks. From the 1st week to the 18th week, the average number of dining activities per student per week is 14.53. However, during the final two weeks, not only do fewer students visit on-campus dining establishments, but the average number of dining activities also decreases to 13 per student, reflecting a decline of approximately 10.5%. If we only consider the 10 weeks when all students attend the CS65 course together, we find that during the 11th week to the 18th week, the average number of on-campus dining activities is 15.37 per week for each student. Comparing that, the average number decreases more than 15% in the final two weeks.

Note that in the 11th week, there is a remarkable surge in both the number of nodes and edges of the student co-occurrence network. This significant increase leads both the number of nodes and the number of edges surpass those of any week within the initial 10 weeks. This advantage continues until the end of the project. A possible reason for this observation is that the students all selected the CS65 course during the latter half of the 20 weeks, and the same course and required course tasks offer them the opportunity to become acquainted with each other and engage in interactions within the class. Then from the perspective of fostering interpersonal relationships, as the students become increasingly acquainted with each other, there inherently arises a desire to engage in communication outside the class. The potential extracurricular interaction venues include on-campus dining establishments. The analysis of dining data provides evidence to support this notion. By comparing the average degree of the student co-occurrence networks between the initial 10 weeks and the latter 10 weeks, we observe a significant increase from 7.47 to 9.70, that is to say, under the premise of attending one shared course, the frequency of contacts among students at the on-campus dining establishments alone increases by roughly 30 percent.

The number of edges of the student co-occurrence networks has a spike in the 13th week and rapidly falls back to the previous level (about 60% of the peak) in the 14th week. The information from the StudentLife project shows that week 13 is close to the deadline for assignments in the CS65 course (see the shade in Figure 1A). As a requirement for earning course credit, students naturally find themselves inclined to increase their communication frequency when they have shared homework assignments. This heightened level of communication serves two purposes. First, sufficient communication enhances the quality of individuals' assignments. Second, assignments in programming class often involve a certain collaborative work, which further promotes communication among students. After the assignment deadline, the extracurricular communication frequency among students naturally reverts to the original range.

In addition to analyzing the network size, we also explore the changes of the student co-occurrence networks over time by examining the clustering coefficients. The clustering coefficient [30] describes the probability of interconnections between the neighbors of a node in a network. In social networks, the clustering coefficient represents the average level of interconnectedness within the immediate circle of friends surrounding an individual. The changes in the clustering coefficients of the student co-occurrence networks are shown in

Figure 1B. It can be seen that the clustering coefficients (solid circles) of the student co-occurrence networks exhibit fluctuations within the approximate range of (0.4, 0.6) throughout the entire 20-week period. There, the clustering coefficients reach the peak value (0.776) in the 13th week, which is also believed to be attributed to the sudden surge in interactions among students, caused by the assignment deadline in the CS65 course.

When compared to random networks with the same degree sequences, is the clustering of the real student co-occurrence networks stronger or weaker? We answer this question by employing the null model. The null model is often applied to evaluate the non-trivial features of certain indicators in complex systems. It can provide the random distribution of other features as their evaluation baseline while keeping some pre-set features unchanged [31-33]. We generate null networks with the same degree sequences as the actual student co-occurrence networks. The implementation strategy is to randomly disconnect edges and crossly reconnect the endpoints. The detailed steps are as follows. (1) Two edges without endpoints overlapped (e.g., $e(u_1, u_2)$ and $e(u_3, u_4)$) are randomly selected. (2) We then check if there is an existing edge between the two pairs of endpoints that are desired for cross-reconnection (e.g., $u_1$ and $u_4$; $u_2$ and $u_3$). If no such edge exists, we disconnect the two original edges and generate two new edges, which is named as one round of edge cross-reconnection. If there is at least one such edge, we return to step (1) and randomly re-select two edges. By repeating steps (1) and (2) for $10|E|$ times (that is, 10 times the number of edges of the initial network), we finally get a null network corresponding to the actual co-occurrence network.

To eliminate the random errors, a total of 100 independent null networks are generated. Then, the clustering coefficients of the 100 null model networks are averaged as the final baseline of the null model (the hollow squares in Figure 1B). The comparison in Figure 1B shows that except for the final week, the clustering coefficients of the actual student co-occurrence networks (solid circles) are always significantly larger than that of the corresponding null networks (hollow squares), suggesting there is a cooperative interpersonal relationship among the students.

## 4. Association analysis

The flourishing trait plays a vital role in interpersonal relationships. Early studies have shown that the flourishing trait is positively correlated with individuals' self-love and pro-sociality [27], and high-quality social relationships and close connections can promote the flourishing of interpersonal relationships [34]. Generally, it is intuitive to suppose that the core nodes within the student co-occurrence networks possess strong responsibility and affinity. These two factors are often instrumental in establishing rapid connections with others and assisting the connection establishments between others. Therefore, it is reasonable to assume that the core nodes in the student co-occurrence networks have high satisfaction in the self-perception of interpersonal relationships, which can enhance their level of flourishing. We verify this by analyzing the correlation between the node centrality within the student co-occurrence networks and the increment of test score of the flourishing scale.

To quantify the importance of students in the co-occurrence networks, three most applied centrality metrics are employed to assess the importance of each node within the student co-occurrence networks, saying degree centrality (DC) [35], closeness centrality (CC) [36], and betweenness centrality (BC) [37]. (see more centralities in [38])

The students participated in the flourishing scale survey twice, once before the 11th week and the other after the 20th week. As a result, our assessment focuses solely on examining the association between the cumulative student co-occurrence networks and the flourishing trait during these 10 weeks. For this purpose, we construct cumulative student co-occurrence network $G_a^W$, which represents the co-occurrence network corresponding to the accumulated dining data from the 11th week to the $W$th week, where $W$=11, 12, $\cdots$, 20. The number of students with both dining data and flourishing scale survey data is 30. The statistics of the 30 students on the flourishing scale survey are shown as follows. The mean and standard deviation are separately 44.23 and 5.96 in the first test ($F_1$), while the two are 43.87 and 8.24 in the second test ($F_2$). The score difference in the two tests is denoted as $\Delta F = F_2 - F_1$. As the relationship between the centrality of each node in the cumulative co-occurrence networks and the flourishing trait is generally not linear, the well-known Spearman's rank correlation coefficient [39] is applied to measure the correlation strength between them.

Table 1. Correlation between the node centrality of the cumulative student co-occurrence networks and the flourishing scale. The Spearman's rank correlation coefficient is applied to quantify the correlation strength. Abbreviations W, F, and NC stand for Week, Flourishing trait, and Node Centrality, respectively. Significance levels: \*$P$<0.1, \*\*$P$<0.05, \*\*\*$P$<0.01.

| Fl \ NC | DC | | | CC | | | BC | | |
|---|---|---|---|---|---|---|---|---|---|
| W | $F_1$ | $F_2$ | $\Delta F$ | $F_1$ | $F_2$ | $\Delta F$ | $F_1$ | $F_2$ | $\Delta F$ |
| 11 | 0.169 | 0.326* | 0.234 | 0.198 | 0.340* | 0.207 | −0.046 | 0.045 | 0.115 |
| 12 | 0.115 | 0.280 | 0.227 | 0.137 | 0.287 | 0.216 | 0.064 | 0.143 | 0.093 |
| 13 | 0.122 | 0.315* | 0.377** | 0.122 | 0.310 | 0.375** | 0.159 | 0.266 | 0.306 |
| 14 | 0.063 | 0.237 | 0.361** | 0.06 | 0.236 | 0.360* | −0.066 | 0.080 | 0.296 |
| 15 | −0.093 | 0.113 | 0.390** | −0.096 | 0.107 | 0.386** | −0.195 | 0.070 | 0.398** |
| 16 | −0.161 | 0.047 | 0.379** | −0.161 | 0.047 | 0.379** | −0.177 | 0.073 | 0.406** |
| 17 | −0.081 | 0.121 | 0.389** | −0.081 | 0.121 | 0.389** | −0.125 | 0.139 | 0.441** |
| 18 | −0.061 | 0.151 | 0.393** | −0.061 | 0.151 | 0.393** | −0.045 | 0.094 | 0.327* |
| 19 | −0.068 | 0.185 | 0.447** | −0.068 | 0.185 | 0.447** | −0.076 | 0.072 | 0.356* |
| 20 | −0.088 | 0.208 | 0.491*** | −0.088 | 0.208 | 0.491*** | −0.065 | 0.114 | 0.407** |

The results of the correlation analysis are shown in Table 1. The results indicate that there is no significant correlation between the node centrality and neither of the two flourishing scale tests. At the same time, there is a significant and positive correlation between the former and the score difference in the two flourishing scale tests. It is probably because one's centrality in the social network will increase with the gradual accumulation of interactions with other students, leading to a change in his/her flourishing state. In particular, there is no significant correlation between node centrality and score difference in the initial two weeks. However, the correlation becomes significant from the 13th week and persisting until the 20th week, the end of the StudentLife project. This observation suggests that the establishment and development of interpersonal relationships gradually lead to individual differences in node centrality. As a result, students who are located at the central region of the cumulative co-occurrence networks become more optimistic and confident, and thus we can observe a consistent and positive correlation between the centrality of each node and the corresponding score difference. Moreover, it is found that the strength of the correlation exhibits an overall upward trend, suggesting that the establishment and improvement of interpersonal relationships have accumulative effects on individuals' flourishing traits. These findings indicate that the state of one's flourishing trait at a certain moment is not related to his/her centrality in social networks. Instead, the level of interaction and communication with others in the offline social networks during this period is related to the change in his/her flourishing trait.

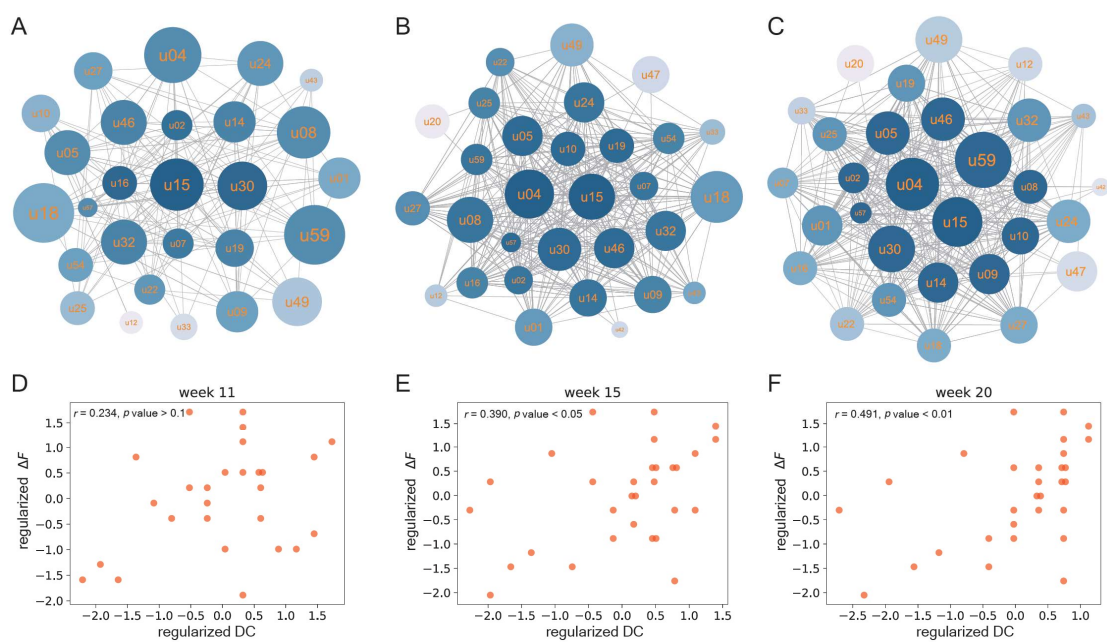

Figure 2. The relationship between degree centrality and the score difference in the two tests of flourishing trait. (A–C) Illustration of cumulative co-occurrence networks $G_a^{11}$ (A), $G_a^{15}$ (B), and $G_a^{20}$ (C). The network layout adopts a "core-periphery" structure. The closer the node is to the center, the larger the degree centrality is. The colors of the nodes represent the intensity of degree centrality. The deeper the color is, the higher the degree centrality is. The sizes of the nodes mean the relative magnitudes of the score differences in the two flourishing scale tests. The larger the node is, the more enhancement the node gains. Since some students experience a decline in the second flourishing scale test, resulting in negative score differences. Therefore, the relative ranking is applied to depict the differences in the change of the flourishing traits among students. (D-F) The scatter plots of the relationships between regularized degree centrality and regularized score difference in the two flourishing scale tests, corresponding to cumulative co-occurrence networks $G_a^{11}$ (D), $G_a^{15}$ (E), and $G_a^{20}$ (F).

By employing the "core-periphery" layout, Figure 2 visualizes the networks $G_a^{11}$, $G_a^{15}$, and $G_a^{20}$. It can be observed that in the cumulative co-occurrence network of the 11th week, several nodes with significant improvement in their flourishing traits stay at the outer place of the networks. At that moment, there is no clear correlation between node centrality and score difference. For ease of comparison across datasets, here we first convert both the centrality and the score difference into relative rankings and then regularize these two sets of relative rankings by leveraging the $Z$-score [40]. Taking the degree centrality as an example, for any student $u$ in the sample, the regularized degree centrality is shown as

$$d'_u = \frac{d_u - \langle d \rangle}{\sigma_d}, \qquad (2)$$

where $d_u$ denotes the degree centrality of student $u$, and $\langle d \rangle$ and $\sigma_d$ respectively represent the mean and standard deviation of degree centrality for all considered students. Similar operations can be used to obtain the regularized score difference in the two flourishing scale tests. Figure 2D shows the scatter plot of relationship between regularized degree centrality and the regularized score difference in the two flourishing scale tests in the 11th week. The Spearman's rank correlation coefficient reveals that there is no statistical significance at that moment.

As time goes on, in the 15th week (Figure 2B), most of the nodes with large improvements in the flourishing trait locate around the center of the network $G_a^{15}$, while only a few nodes with large improvements in the flourishing trait remain stay at the periphery. At that point, a correlation has emerged between the degree centrality of nodes and the score difference. Nodes that are closer to the center of the network have more likely to exhibit significant improvement in the flourishing trait, and vice versa. The corresponding scatter plot (Figure 2E) and Spearman's rank correlation coefficient confirm the significance of the correlation.

For the cumulative co-occurrence network in the 20th week (Figure 2C), it can be observed apparently that students with the highest improvement in the flourishing trait are all located at the central region of the network, suggesting a significant and positive correlation between the degree centrality of nodes and the improvement of the flourishing trait.

The results by leveraging closeness centrality and betweenness centrality for centrality quantification are similar to that of the degree centrality, as shown in Table 1. In addition to correlation analysis, we also utilize the "core-periphery" layout to visualize the cumulative student co-occurrence networks on the closeness centrality measure and betweenness centrality measure. Similar to the case of degree centrality, we observe a progress of the correlation between node centrality and the score difference, from non-significant to significant. Moreover, there is a trend that nodes with large improvements in the flourishing trait gradually moving to the network center.

## Conclusion and discussion

This study constructs student co-occurrence networks based on students' on-campus dining activities and analyzes the topology changes over 20 weeks, finding that there is a correlation between the node centrality and the flourishing trait. Specifically, the node centrality experiences a trend from being unrelated to significantly and positively correlated to the score difference in the two flourishing scale tests. Similar results are observed under three different centrality measures, indicating the robustness of the aforementioned conclusion.

For an individual in an offline social network, there is no direct connection between his/her position in the network and his/her flourishing state at a certain moment. However, with the accumulation of interactions with others, the centrality of that individual within the network strengthens gradually, thereby affecting his/her personal flourishing trait and causing the change in his/her flourishing state. This inspires us to encourage individuals who are less connected in the social network to reinforce their interpersonal interactions with others. After interactions and communications accumulate to a certain degree, these individuals staying in the peripheral region will naturally gravitate toward the central area of the network. Moreover, the increase in the quality of social interactions can strengthen an individual's self-perception of optimism and confidence, which further results in the improvement of their mental well-being levels. For educational administrators, this may provide them a reference in guiding students to maintain a positive mental health status in the future, for example, by encouraging them to engage more in social interactions with peers.

Although our work helps to deepen the understanding of the relationship between students' offline social networks and their psychological traits, the interpretation of the research findings should not go beyond the limitations of the data and analysis. Firstly, there are only 31 students in the sample, which is relatively small, resulting in insufficient representativeness of the sample. Secondly, introducing a time threshold $T$ to determine whether co-occurrence happens makes it impossible to rule out the possibility of the "fake edges" noise generated by accidental spatial-temporal overlaps in dining activities. Moreover, social activities outside of dining establishments may also affect students' offline social networks, but it has not been considered in this study. Furthermore, the personal information of students is unknown, making it impossible to further explore whether the heterogeneity of students (for example, gender and major) has an impact on the node centrality of social networks and the personal flourishing trait. Although there is sufficient statistical significance in the correlation analysis, we still look forward to the opportunity to test the universality of the findings of this study on a larger student group from diverse educational and cultural backgrounds in the future, and to explore the possibility of extending it to other mental health traits.

## Acknowledgements

This work is partially supported by the National Natural Science Foundation of China (Grant No.42361144718) and the Ministry of Education of Humanities and Social Science Project, China (Grant No. 21JZD055). The funders had no role in the study design, data collection, analysis, decision to publish, or preparation of the manuscript.